\documentstyle[emulateapj]{article}

\lefthead{Natarajan \& Pringle}
\righthead{Alignment of disk and black hole spins in AGN}

\eqsecnum
\def\beq{\begin{equation}}
\def\eeq{\end{equation}}
\def\simge{\mathrel{%
   \rlap{\raise 0.511ex \hbox{$>$}}{\lower 0.511ex 
\hbox{$\sim$}}}}
\def\simle{\mathrel{
   \rlap{\raise 0.511ex \hbox{$<$}}{\lower 0.511ex 
\hbox{$\sim$}}}}
\def\ref{\reference}

\begin{document}
\title{The alignment of disk and black hole spins in active galactic nuclei}
\author{Priyamvada Natarajan$^{1,2}$, and J. E. Pringle$^{1,3}$}
\affil{$^1$Isaac Newton Institute for Mathematical Sciences, 20 Clarkson Road, Cambridge , U. K.}
\affil{$^2$Canadian Institute for Theoretical Astrophysics, McLennan Labs., 
60 St. George Street, Toronto M5S A1, Canada; \\ priya@cita.utoronto.ca} 
\affil{$^3$Institute of Astronomy, Madingley Road, Cambridge CB3 0HA, U. K.
; \\ jep@ast.cam.ac.uk}

\begin{abstract}

The inner parts of an accretion disk around a spinning black hole are
forced to align with the spin of the hole by the Bardeen-Petterson
effect. Assuming that any jet produced by such a system is aligned
with the angular momentum of either the hole or the inner disk, this
can, in principle provide a mechanism for producing steady jets in AGN
whose direction is independent of the angular momentum of the
accreted material. However, the torque which aligns the inner disk
with the hole, also, by Newton's third law, tends to align the spin of
the hole with the outer accretion disk. In this letter, we calculate
this alignment timescale, $t_{\rm{align}}$ for a black hole powering
an AGN, and show that it is relatively short. This timescale is
typically much less than the derived ages for
jets in  radio loud AGN, and implies that the jet directions are not
in general controlled by the spin of the black hole. We speculate that
the jet directions are most likely controlled either by the angular
momentum of the accreted material or by the gravitational potential of
the host galaxy.

\end{abstract}

\keywords{accretion, accretion disks $-$ black hole physics:   
$-$ galaxies: nuclei, active, jets}

\section{Introduction}

The classic double radio sources, the FRIIs (Faranoff \& Riley
1974), are thought to be powered by 
narrow beams or jets of energy
emanating from a nuclear black hole in the center of the host
galaxy (Rees, 1971, 1984; Longair, Ryle \& Scheuer 1973). In many, though 
not all, of these sources, the spatial direction of the jets has
remained unchanged throughout the lifetime of the radio source. In
the most luminous sources the lengths of the jets reach up to scales of
around a Mpc, and the source lifetimes are as long as around $10^8$
years (Alexander \& Leahy 1987, Liu, Pooley \& Riley 1992). The 
degree of collimation of the beams, often of the order of 0.1 radian or 
less, implies that whatever physical mechanism gives rise to the 
beam stability is unchanging in the direction it defines on 
comparable timescales, that is of at least $10^9$ years. In this letter, 
we address the question of what might give rise to such stability.

This problem was considered by Rees (1978) who came to the conclusion
that the most likely cause of the stability of the jet directions is
due to the fact that it is determined by the spin of the nuclear black 
hole. Any accretion disk flow which might power the nuclear activity is 
aligned with the black hole spin direction by the Bardeen-Petterson 
effect (Bardeen \& Petterson 1975) out to a disk radius 
of $R_{\rm BP}\,>>\,R_s$, where $R_s$ is 
the Schwarzschild radius. Thus, if the jet is disk
powered, the jet will be strictly aligned with the spin of the
hole. Furthermore, this idea also fits into the current
theoretical paradigm in which jets are powered directly by
the spin energy of the hole (Blandford 1991, and references therein;
Rawlings \& Saunders 1991), although this paradigm is open to debate (see, for 
example, Livio, Ogilvie \& Pringle 1998). 
Rees (1978) pointed out however, that the couple
exerted by the black hole as it aligns the disk with the spin of the
hole, has by Newton's third law also the effect of aligning the hole
with the spin of the disk. He estimated the timescale for this
alignment to occur by assuming that each accreted mass element brings 
with it to the hole angular momentum corresponding to its orbital angular
momentum at the Bardeen-Petterson radius $R_{\rm BP}$. He suggested that 
this alignment timescale is given by,
\beq 
t_a\,\sim\,{{\frac{M}{\dot M}}{\frac{J}{J_{\rm max}}}({\frac{R_s}{R_{\rm BP}}})^{1/2}}, 
\eeq
where ${M/\dot M}$ is the accretion time-scale $t_{\rm acc}$, and 
${J/J_{\rm max}}$ is the ratio of the angular momentum of the black hole 
to the maximal angular momentum of a Kerr black hole. Rees (1978) estimated 
$t_{\rm a}$ to be of the order of $10^8$ years.

However, in recent years there has been considerable theoretical 
progress in our understanding of how a warped (or non-planar) 
accretion disk communicates the warp and evolves in time. 
This physical process 
has a direct bearing on the black hole/accretion disk alignment 
timescale. We show below, that, using current theories, the alignment 
timescale is considerably shorter than the ages of the radio sources 
inferred from spectral ageing models fitted to the observations, and is
consequently also less than the timescale on which the jet direction
changes. 

\section{The alignment process}

The timescale for disk/hole alignment depends directly on the
timescale on which an accretion disk can transfer a warp in the radial
direction. In an accretion disk the component of angular momentum
parallel to the spin of the disk is transferred at radius $R$ in the
disk in a diffusive manner on a timescale $t_R\,\sim\,R^2/\nu_1$, 
where $\nu_1$ is the usual disk kinematic viscosity (Pringle 1981; 
Frank, King \& Raine 1992). Using the dimensionless viscosity 
parameter $\alpha$ (Shakura \& Sunyaev 1973), this timescale can 
be written approximately as, 
\beq
t_R\,=\,\Omega^{-1}\,({\frac{R}{H}})^2\,\alpha^{-1}, 
\eeq
where $H$ is the local disk semi-thickness and $\Omega$ the local angular
velocity. It was originally assumed (Bardeen \& Petterson 1975; 
Rees 1978) that the component of the disk
angular momentum lying in the plane of the disk (that is, the warp) is
transferred radially on a similar timescale. However, it was
discovered by Papaloizou \& Pringle (1983) that consideration of the
propagation of disk warp must necessarily take into account the internal
hydrodynamics of the disk itself. In the regime in which ${H/R}\,<\,{\alpha}
\,<<1$, and in which the disk is close to being Keplerian, they found (see
also Kumar \& Pringle 1985) that the disk behaviour is somewhat
complicated, but that to a first approximation the component of
angular momentum in the disk plane is transferred within the disk on a
timescale of order $R^2/\nu_2$, where $\nu_2/\nu_1 = 1/2 \alpha^2$ (assuming 
that $\alpha\,<<\,1$). 
Thus, the relevant timescale for communication of the disk warp is, 
\beq
{t_{\rm warp}}\,\sim\,{\alpha^2}\,{t_R}\,\,<<\,t_R. 
\eeq
The original calculations by Papaloizou \& Pringle (1983) were carried
out using Eulerian linear perturbation theory about an initially flat
disk, and so were formally only valid for disk warp angles, $\beta$, much
less than the disk opening angle $H/R$. For AGN disks for which $H/R\,\sim\, 10^{-2} --
10^{-3}$ (see below), this is somewhat limiting. Recent work by Ogilvie
(1998a,b)  however, has shown that similar conclusions remain valid for 
warps of significant amplitude. These results have a considerable 
effect on the so-called Bardeen-Petterson radius, $R_{\rm BP}$, the 
radius out to which the disk is aligned with the spin of the hole, as 
well as on the hole/disk alignment timescale. Since the disk turns out 
to be far more efficient at transferring warp in the radial direction than 
the initial estimates, which had ignored the internal disk hydrodynamics, 
it follows that both $R_{\rm BP}$ (Kumar \& Pringle 1985) and the 
alignment timescale are much smaller than was originally thought.

\subsection{The alignment radius}

The timescale on which a misaligned black hole aligns with its disk
and the radius out to which the alignment occurs have been calculated
by Scheuer \& Feiler (1996). Writing the Lense-Thirring precession rate
in the disk as $\Omega_{\rm LT}\,=\,\omega_p/R^3$, they find that the radius out to which
the disk is aligned with the spin of the hole is given simply as
the radius at which the timescale for radial diffusion of the warp,
$t_{\rm warp}$, is of the order of the local Lense-Thirring 
precession timescale $\Omega_{\rm LT}^{-1}$. Equating these we obtain, 
\beq 
{R_{\rm warp}}\,\sim\,{\omega_p/\nu_2}.
\eeq
where $\omega_p = 2 G J / c^2$, the angular momentum of the 
hole, J, is given by $J\,=\,a\,c\,M\,(G\,M/c^2)$,
$M$ is the mass of the hole, and $a$ ($0\,<\,a\,<\,1$) the dimensionless spin
parameter. Using these expressions, together with the fact that
$\nu_2/\nu_1\,=\,1/{2\,\alpha^2}$, and writing $\nu_1\,=\,\alpha\,H^2\,\Omega$, 
we find that $R_{\rm warp}$ may be written as,
\beq
{\frac{R_{\rm warp}}{R_s}}\,=\,2\,a\,\alpha\,({\frac{R}{H}})^2\,
({\frac{R\,\Omega}{c}}),
\eeq
where $R_s\,=\,2\,G\,M/c^2$ is the Schwarzschild radius.
Taking account of the fact that far from the hole, 
\beq
{\frac{R\,\Omega}{c}}\,=\,{\frac{1}{\sqrt 2}}\,({\frac{R_s}{R}})^{1/2},
\eeq
we find that,
\beq
{\frac{R_{\rm warp}}{R_s}}\,\nonumber=\,{2^{1/3}}\,(a\,\alpha)^{2/3}\,({\frac{R}{H}})^{4/3}.
\eeq

To proceed further we need a model for the AGN disk at the relevant
radii. We make use of the AGN disk models computed by
Collin-Souffrin \& Dumont (1990) from which in the relevant range of
radii we find that,
\begin{eqnarray}
{\frac{H}{R}}\,=\,7.1\,\times\,{10^{-3}}\,({\frac{\alpha}{0.03}})^{-1/10}\,
({\frac{L}{0.1\,L_E}})^{1/5} \\ \nonumber
\times M_8^{-1/10}\, ({\frac{\epsilon}{0.3}})^{-1/5}\,({\frac{R}{R_s}})^{1/20}.
\end{eqnarray}
Here $\epsilon$ is the efficiency of the accretion process defined as,
$\epsilon\,=\,{L/{\dot M c^2}}$, and $L_{\rm E}$ is the Eddington
luminosity, $L_{\rm E}\,=\,1.4 \times 10^{46}\,M_8$ erg s$^{-1}$,
where $M_8$ is the black hole mass in units of $10^8\,M_\odot$.
Throughout this letter we 
shall take $\alpha\,=\,0.03$ and $L\,=\,0.1\,L_E$ to represent the typically 
expected standard values.

Using this we find that,
\beq
{\frac{R_{\rm
warp}}{R_s}}\,\nonumber=\,66\,a^{5/8}\,({\frac{\alpha}{0.03}})^{3/4}\,({\frac{L}{0.1\,L_E}})^{-1/4}\,
{M_8}^{1/8}\,({\frac{\epsilon}{0.3}})^{1/4}.
\eeq
This expression is valid provided that
\beq
{\frac{L}{L_{\rm E}}}\,>\,2.0\,\times\,{10^{-2}}\,a^{9/10}\,({\frac{\alpha}{0.03}})^{7/5}\,
({\frac{\epsilon}{0.3}})\,M_8^{1/2}.
\eeq
We note that although Scheuer and Feiler (1995) used a simplified set
of evolution equations (Pringle 1992) which take into account the
difference between $\nu_1$ and $\nu_2$, but do not take the full effects of
internal disk hydrodynamics into account, their estimates of the
alignment radius are in substantial agreement with the full
calculations of Kumar \& Pringle (1985) for values of $\alpha\,\simle\,0.3$. 

\subsection{The alignment timescale}

Scheuer \& Feiler (1996) find that the effect of the disk on the black
hole is to force the spin axis of the hole to precess and to align
with the disk. Both precession and alignment take place on the
same timescale which is given by,
\beq
{t_{\rm align}}\,\sim\,({\frac{J}{J_d}})\,\Omega_{\rm LT}^{-1}, 
\eeq 
where $J_d$ is the angular momentum of the disk within the warp radius
$R_{\rm warp}$, and $\Omega_{\rm LT}$ is the Lense-Thirring angular velocity 
also evaluated at $R_{\rm warp}$. It should be noted that the transfer of 
angular momentum between the hole and the disk does not depend in any 
way on the disk being an accretion disk (i.e. it is independent of $\dot M$).
Therefore, the above formula is valid provided that the
disk is able to transfer warp, which it may do purely by diffusion if $\alpha 
> H/R$, or by warp waves (Lubow \& Pringle 1993, Papaloizou \& Lin 1995,
Nelson \& Papaloizou 1998) if $\alpha\,<\,H/R$, and provided that the disk
can act as an adequate sink of angular momentum. In the viscous case, 
($\alpha\,>\,H/R$) 
relevant to AGN disks, this timescale may be rewritten (Scheuer \& Feiler 1995) as,
\beq
t_{\rm align}\,\sim\,({\frac{a\,c\,M}{2\,G\,\nu_2}})^{1/2}\,{\frac{1}{\pi\,\Sigma}},
\eeq
where $\Sigma$ is the surface density of the disk at the warp radius
$R_{\rm warp}$. For a steady accretion disk far from the centre, it can be shown
that (Pringle 1981),
\beq
\Sigma\,=\,{\frac{\dot M}{3\,\pi\,\nu_1}}.
\eeq
Thus, for a steadily accreting disk, we may write,
\beq
{\frac{t_{\rm align}}{t_{\rm acc}}}\,=\,1.9\,\times\,{10^{-2}}\,{a^{1/2}}\,
({\frac{\alpha}{0.03}})^{3/2}\,(\frac{H}{R})\,
(\frac{R_{\rm warp}}{R_s})^{1/4},
\eeq
where the accretion timescale $t_{\rm acc}$ is defined as 
$t_{\rm acc}\,=\,M/\dot M$.

Alternatively, for a steady disk, the accretion timescale can be
written in terms of the Salpeter time $t_S$, which is the growth
timescale for the black hole if it is accreting at a rate corresponding 
to the limiting Eddington luminosity $L_E$.
Thus, 
\beq
t_{\rm acc}\,=\,{t_S}\,({\frac{L}{L_E}})^{-1},
\eeq
where, $t_S\,=\,1.2\,\times\,10^8\,({\epsilon}/{0.3})$ years.

Making use of equations 2-7, 2-8, 2-13 and 2-14, we find that
\begin{eqnarray}
{\frac{t_{\rm align}}{t_S}}\,=\,4.7\,\times\,{10^{-3}}\,a^{11/16}\,
({\frac{\alpha}{0.03}})^{13/8}\,(\frac{L}{0.1\,L_E})^{-7/8} \\ \nonumber 
\times M_8^{-1/16}\,({\frac{\epsilon}{0.3}})^{-1/8},
\end{eqnarray}
or equivalently,
\begin{eqnarray}
t_{\rm align}\,=\,5.6\,\times\,10^5\,\,a^{11/16}\,
({\frac{\alpha}{0.03}})^{13/8}\,(\frac{L}{0.1\,L_E})^{-7/8} \\ \nonumber
\times M_8^{-1/16}\,({\frac{\epsilon}{0.3}})^{7/8}\,
{\rm years}. 
\end{eqnarray}

\section{Discussion}

We have shown that our current understanding of the time-evolution of
warps in accretion disks leads to the conclusion that a disk which is
misaligned with the spin of a central black hole, and whose inner
regions are therefore aligned with the spin of the hole by the
Bardeen-Petterson effect, brings the spin vector of the hole into
alignment with the spin vector of the outer disk on a timescale which
is much shorter than had previously been realised. Thus, the idea that
the maintenance of the jet directions in extended double radio sources
for timescales of up to of order $10^9$ years is due to the `flywheel'
effect of the central spinning black hole is no longer a tenable one.
This should perhaps in any case not be too surprising since for some
kind of flywheel mechanism to work one would normally select as the
flywheel an object with as large a moment of inertia as
possible. Taking cognizance of this, the idea that the smallest object
in the system might be the flywheel seems in retrospect counter-intuitive. 
Thus a more likely identification as the flywheel might be the accretion disk
itself. For example, the radius at which black hole and disk angular
momenta are equal for the Collin-Souffrin \& Dumont (1990) AGN models is
\beq
{\frac{R}{R_s}}\,\sim\,1.4\,\times\,{10^4}\,a\,{M_8}^{-1}\,({\frac{\alpha}{0.03}})\,(\frac{\epsilon}{0.3})\,(\frac{L}{0.1\,L_{\rm
E}})^{-1}.
\eeq
In this picture, the constancy of directionality of the jets is due to
the fact that the accretion event was of a single gas rich
object. Alternatively, one might consider using the host galaxy as
the flywheel, if the galactic potential is such that any accreted gas
would soon start to orbit in some preferred plane.

We should note that although the overall conclusions are unlikely to
change, the details of the alignment timescales and disk warp
structure will depend n the specific accretion disk model applicable
in any given case. For illustration, in this paper, we have used
the models by Collin-Souffrin \& Dumont (1990). However, it should be
borne in mind, that even these models strictly need some amendment
to take account of assymetric heating of the disk because it is
twisted and of additional internal heating due to the enhanced $\nu_2$
type of dissipation brought on by the twist. In addition, it may be that
the accretion disk  structure is quite different from the standard thin
disk structure envisaged by Collin-Souffrin \& Dumont (1990). For 
example, if the flow onto the hole is in the form of an advection-dominated
flow (ADAF), for which $\alpha\,\sim\,H/R\,\sim\,1$ , (Narayan \& Yi, 1995)
 then from equation 2.4 we see that we expect $R_{\rm warp}/R_s\,\sim\,1$, 
together with a consequently reduced alignment timescale. All these issues
will need to be addressed in further work.

From an observational point of view, if the jet direction is
controlled solely by the black hole spin, then, because the
Bardeen-Petterson radius is typically at radii too small to be observed
directly, we would not necessarily expect to observe a correlation
between the angular momentum vector of gas in the host galaxy, and the
directions of the jets. In this context, we note that in a survey of
nearby radio loud early type galaxies with HST, van Dokkum and Franx
(1995) find that the major axes of the dust disks seen in the inner
regions ($\simle$ 250 pc) are aligned perpendicular to the arc second radio
structures in the nuclei. This might be taken as {\it prima facie} 
evidence that the direction of the radio jets in these objects is
determined by the galactic potential and/or by the orbital angular
momentum of the gas rich intruder which presumably produced both the
dust disks and the nuclear activity. We also note that for this to
occur, and indeed for jets to be able to display long term directional
stability, then the radiation driven (perhaps wind enhanced) warping
mechanism discussed by Pringle (1996,1997) must be unable to operate in
these systems. This can occur either because the mechanism is not
powerful enough to operate in galactic nuclei (it seems likely that
the effect may well need wind enhancement in this case since it is
severely reduced for values of $\nu_2/\nu_1\,>>\,1$) or because the warping
is stabilised by Lense-Thirring precession driven by a sufficiently
rapidly spinning central black hole (Pringle 1997; Maloney, private
communication). We note that the argument here differs somewhat from
that of Wilson \& Colbert (1995) who assume that a rapidly spinning
black hole is necessary to produce mechanical jet energy. The present
line of reasoning would suggest that while we already know from the
 protostellar case that the presence of a black hole (spinning or not)
is not necessary for the production of a jet (see, for example, Pringle
1991), it is the spin of the hole which is required to suppress the
warping instability (which only operates effectively in disks around
compact objects) and so produce a jet with some directional stability.

In contrast to this, however, is the indication that in the nuclei of
Seyfert (spiral) galaxies, the direction of the arc second scale radio
structure (presumed to be the inner radio jets) is independent of the
spin axis of the disk of the spiral galaxy, and indeed appears to be oriented
completely at random in space (Clarke, Kinney \& Pringle 1998). If our
estimates of the hole/disk alignment are correct, then there are two
obvious possibilities to explain this. First, the radiation driven
warping instability may be able to operate in these nuclei, presumably
because the black hole spin is much reduced. We note that if the inner
disk direction is indeed randomly directed over large time then the
net angular momentum accreted by the hole from the disk is much
reduced (Moderski \& Sikora 1996a), and that at the low accretion rates
relevent to Seyfert nuclei it may be possible to spin the black hole
down (Moderski \& Sikora 1996b). If this is so, then in order to
produce rapidly spinning black holes in galactic nuclei then it may be
necessary, following Wilson \& Colbert (1995) to appeal to black hole
mergers. And, second, the gas being fed to the hole may come, not from
the disk of the galaxy, but from some other source, such as a small gas
rich intruder on a plunging orbit to the nucleus. In any case to test
these ideas further it will be important to observe the dynamics of gas on
small scales (10-100 parsec) around Seyfert nuclei.

\section*{Acknowledgments}

We acknowledge the Isaac Newton Institute for Mathematical Sciences
and specifically the organizers and participants of the workshop on the
Dynamics of Astrophysical Discs for useful discussions and providing 
a vibrant atmosphere.

\vfill\eject

\end{document}